\documentstyle[12pt]{article}

\def\thefootnote{\fnsymbol{footnote}}
\renewcommand{\thefootnote}{\alph{footnote}}

\newcommand{\rref}[1]{(\ref{#1})}
\newcommand{\beqn}{\begin{equation}}
\newcommand{\eeqn}{\end{equation}}
\newcommand{\beqarr}{\begin{eqnarray}}
\newcommand{\eeqarr}{\end{eqnarray}}

\newcommand{\matc}{\begin{array}{c}}
\newcommand{\matcc}{\begin{array}{cc}}
\newcommand{\matccc}{\begin{array}{ccc}}
\newcommand{\matcccc}{\begin{array}{cccc}}
\newcommand{\emat}{\end{array}}
\newcommand{\df}{\stackrel{\rm def}{=}}
\newcommand{\SOS}{SO(3,3\,|{\bf Z})}
\newcommand{\SOdd}{SO(d,d\,|{\bf Z})}
\newcommand{\CTPD}{({\cal C}\Theta+{\cal D})}

\begin{document}

\begin{titlepage}

November 1998         \hfill
\begin{center}
\hfill    UCB-PTH-98/56 \\
\hfill    LBNL-42548	 \\

\vskip .15in
\renewcommand{\thefootnote}{\fnsymbol{footnote}}
{\large \bf T-Duality and Ramond-Ramond Backgrounds\\ 
in the Matrix Model}\footnote{This 
work was supported in
part by the Director, Office of Energy Research, Office of High Energy and 
Nuclear Physics, Division of High Energy Physics of the U.S. Department of 
Energy under Contract DE-AC03-76SF00098 and in part by the National Science
Foundation under grant PHY-95-14797}
\vskip .30in

Daniel Brace\footnote{email address: brace@thwk2.lbl.gov}, 
Bogdan Morariu\footnote{email address: morariu@thsrv.lbl.gov} and 
Bruno Zumino\footnote{email address: zumino@thsrv.lbl.gov}
\vskip .30in

{\em 	Department of Physics  \\
	University of California   \\
				and	\\
	Theoretical Physics Group   \\
	Lawrence Berkeley National Laboratory  \\
	University of California   \\
	Berkeley, California 94720}
\end{center}
\vskip .40in

\begin{abstract}
We investigate T-duality of toroidally compactified
Matrix model with arbitrary Ramond-Ramond backgrounds 
in the framework of noncommutative super Yang-Mills 
gauge theory. 
\end{abstract}
\end{titlepage}

\newpage
\renewcommand{\thepage}{\arabic{page}}

\setcounter{page}{1}
\setcounter{footnote}{0}

% main text is here

\section{Introduction}
\label{Intro}

The original Matrix model conjecture~\cite{BFSS}, which 
required taking the large $n$ limit, was further refined in~\cite{LS} by
Susskind who conjectured that the discrete light-cone quantization
(DLCQ) of M-theory is given by the finite $n$ Matrix model~\cite{CH}.
Upon compactification on a torus $T^d$ there are additional degrees 
of freedom, which correspond to winding states in the auxiliary Type IIA
string theory~\cite{LS,AS}. 
The compactified Matrix model is described by a super 
Yang-Mills gauge theory on the dual torus.
Alternatively, one could also ask directly if compactified 
Type IIA string theory in the  $n$ D0-brane charge sector is
equivalent to
a $U(n)$ super Yang-Mills gauge theory. 
An equivalence between string theory and 
a finite rank gauge theory is only possible in the limit of
decoupling string excitations which requires taking  $\alpha'$ to zero.
To  keep the winding excitations finite this implies we must also take 
the small volume limit.
This is exactly the limit
in which the Hamiltonian of the auxiliary Type IIA string theory
equals the DLCQ Hamiltonian of M-theory~\cite{AS}.

For $d \geq 2$ besides the compactification metric 
there are additional moduli which, in terms of the auxiliary Type IIA
string theory~\cite{AS}, correspond to  the $2$-form of the 
NS-NS (Neveu Schwarz-Neveu Schwarz) sector
and the R-R (Ramond-Ramond) forms. 
In the seminal paper~\cite{CDS} Connes, Douglas and 
Schwarz conjectured
that the  $2$-form of the NS-NS sector corresponds to the 
deformation parameter of a noncommutative super Yang-Mills (NCSYM) 
gauge theory. Further studies followed in~\cite{DH,AS1,AS2,BMZ,HV,DBBM,KS}. 

In this paper we continue our investigation started in~\cite{BMZ} 
of T-duality of the DLCQ of toroidally compactified M-theory 
and its realization in terms of NCSYM gauge theory. 
In~\cite{BMZ} we gave an explicit description of the relationship proposed 
in~\cite{AS1,AS2} between T-duality of string theory  
and the duality of the NCSYM gauge theory
known in the mathematical literature as Morita equivalence~\cite{R1}.
Here we extend the previous results by allowing arbitrary 
R-R backgrounds. 

In general two NCSYM theories are dual to each other if there exists an
element $\Lambda$ of the duality group 
$SO(d,d\,|{\bf Z})$ with the block decomposition\footnote{The 
$SO(d,d\,|{\bf Z})$ subgroup of the T-duality group $O(d,d\,|{\bf Z})$ is
the subgroup that does not exchange Type IIA and IIB string theories.}
\beqn
\Lambda=
\left(
\begin{array}{cc}
{\cal A} & {\cal B} \\
{\cal C} &{\cal  D}
\end{array}
\right),
\label{lamb}
\eeqn
such that their defining parameters
are related as follows
\beqarr 
\bar{\Theta}
&=&
 ({\cal A}\Theta +{\cal B})({\cal C}\Theta+{\cal D})^{-1},
\label{Theta299} \\
\bar{G}^{ij} 
&=& 
({\cal C}\Theta +{\cal D})^{i}_{\,k} 
({\cal C}\Theta +{\cal D})^{j}_{\,\,l} \, G^{kl},
\label{dual}   
\\
\bar{g}_{SYM}^{2}  
&=& 
\sqrt{\,|\det({\cal C}\Theta+{\cal D})|}~g_{SYM}^{2},
\label{gsym}
\\
\bar{\eta} &=& S(\Lambda) \eta,
\label{eta}
\\
\bar{\chi} &=& S(\Lambda) \chi.
\label{Chitrans}
\eeqarr
Here $S(\Lambda)$ denotes the  Weyl spinor representation of $\Lambda$ and
$\eta$ is an integral chiral spinor containing the Chern numbers of the
bundle. For compactification on a three torus $\eta$  
contains the rank of the group and the magnetic fluxes. 

The first four relations can be found in~\cite{AS1,AS2,BMZ}. 
Deriving~\rref{Chitrans} 
in the context of NCSYM gauge theory 
is the main thrust of our paper.
The chiral spinor $\chi$ in~\rref{Chitrans} determines the parameters of the
Chern-Simon type terms which we will add to the NCSYM action. In the auxiliary
Type IIA string theory $\chi$ is closely related to the R-R moduli.

In Section~\ref{SUGRA} we review the transformation properties of 
the R-R moduli under the duality group. The dimensionally reduced action  
of Type IIA supergravity is invariant under the
T-duality group\footnote{The equations of motions are invariant under 
the full U-duality group $E_{d+1(d+1)}$.} $SO(d,d)$. By deriving the 
nonlinear sigma model 
which describes the scalar fields of the supergravity, we can extract the 
transformation properties of the R-R backgrounds under the duality group. 
In particular we will show that appropriately defined fields, which are 
combinations of the R-R forms and the NS-NS two-form,
transform in a spinor representation of the duality group.

In Section~\ref{DUAL}, we identify the Chern-Simon 
parameters of the gauge theory with the R-R moduli. Then, we show that the 
duality transformations relating different NCSYM theories 
can be extended to include these terms. 
In the process we obtain the 
transformations properties of the parameters and show that they coincide 
with the transformations expected from string theory and derived in 
Section~\ref{SUGRA} using the dimensional reduction of $10$-dimensional 
Type IIA supergravity.

Finally in the Appendix we present some results, 
used in the main text, regarding
transformation properties under the T-duality group in the 
limit of small compactification volume and decoupling of string excitations.

After obtaining these results we received an e-print~\cite{HV} containing a
similar proposal for the additional terms in the noncommutative action.

\section{Duality of Seven Dimensional Supergravity}
\label{SUGRA}

Type IIA superstring theory compactified on a d-dimensional torus 
is invariant under the T-duality group $SO(d,d\,|{\bf Z})$. 
The low energy supergravity effective action describing  
this compactification is in fact invariant under the continuous group 
$SO(d,d)$.
This action can be obtained directly from the $10$-dimensional 
Type IIA supergravity by dimensional reduction. 
In this section we are interested in  obtaining
the transformation properties of 
the R-R moduli under the discrete duality group. Since this is a 
subgroup of the corresponding continuous group which is a symmetry of the 
low energy 10-dimensional supergravity action, we can
obtain these transformation properties by analyzing the symmetries of the 
the nonlinear sigma model which describes the dynamics of the scalars
in the supergravity action. 

The NS-NS scalars are described locally
by an $O(d,d)\, / \,O(d)\times O(d)$ nonlinear sigma model.
Taking into account the T-duality group,
the NS-NS  nonlinear sigma model is in fact defined on
\[
  O(d,d\,|{\bf Z})     \setminus 
O(d,d)\, / \,O(d)\times O(d).
\] 
On the 
other hand simple counting arguments suggest that the R-R scalar fields
transform in a chiral spinor representation of the duality group. This 
statement is almost correct except that the fields which 
transform in the spinor representation are some redefined fields 
involving not only the R-R fields but also the NS-NS two form.

The $10$-dimensional supergravity action written in terms of the
string metric is given by 
\[
{\cal S} = \int d^{10}x \sqrt{g}~ e^{-2\phi}~(R +4 (\nabla \phi)^2 -
\frac{1}{2 \cdot 3!} H^2) 
\]
\[
-\int d^{10}x \sqrt{g}~
(\frac{1}{2 \cdot 2!} F^2 + 
\frac{1}{2 \cdot 4!} F'^2) 
\]
\[
-\frac{1}{4}~\int F_{(4)} \wedge F_{(4)} \wedge B + \ldots,
\]
where we have not written the terms containing the fermionic fields. The 
first line contains only NS-NS fields while the second
contains the kinetic terms of the R-R forms. 
The various field strengths are defined as follows
\beqarr
H &=& dB,    \nonumber  \\
F &=& dA_{(1)} ,   \nonumber  \\
F_{(4)} &=& dA_{(3)},    \nonumber  \\
F' &=& F_{(4)} + A_{(1)} \wedge dB,    \nonumber  
\eeqarr
where the subscript indicates the rank of the form.
Note that R-R fields couple to the NS-NS fields 
through the metric and through  the $F'^{2}$ term,
which depends on the antisymmetric
NS-NS two-form. 

Next we perform the dimensional reduction along
coordinates $x^i$ for $i=1,2,3$. 
The massless scalars from the NS sector can be organized in the 
symmetric matrix~\cite{ASen}
\beqn
{\cal M} = 
\left(
\begin{array}{cc}
G^{-1} & -G^{-1}B \\
BG^{-1} & G  -BG^{-1}B
\end{array}
\right).
\label{M}
\eeqn
Note that ${\cal M}$ is also an element of the group $SO(3,3)$.
Using a result from the Appendix, we can obtain the  Weyl spinor 
representation of ${\cal M}$
\[
S({\cal M})= 
\left(
\begin{array}{ccc}
\det G^{-1/2} &&\det G^{-1/2}~b^{T} \\
\det G^{-1/2}~b &&\det G^{1/2}~G^{-1}+ \det G^{-1/2}~b \,b^{T}
\end{array}
\right),
\]
where $b=*B$, and the star denotes the operator which transforms an 
antisymmetric matrix into its dual column matrix. The star
operator always dualizes only with respect to the compactified coordinates.

We obtain additional scalars from the dimensional reduction of
R-R forms. As mentioned above these 
fields do not have simple transformation properties under 
the T-duality  group but we can define the following odd rank forms
\beqarr
C_{(1)} &=& A_{(1)},   \label{Cfields}  \\
C_{(3)} &=& A_{(3)}  - A_{(1)}\wedge  B,  \nonumber
\eeqarr 
and organize them in a column matrix which, as we will see shortly,
transforms in a chiral spinor 
representation of the duality group
\[
\chi=\left(
\begin{array}{c}
C_{123} \\
C_1  \\
C_2  \\
C_3  
\end{array}
\right).
\]

The other fields can also be organized in representations of the 
duality group such that the action obtained by dimensional 
reduction from $10$-dimensional supergravity is explicitly invariant.
The six vectors obtained from the dimensional reduction of 
NS-NS fields transform in the fundamental representation
while the $7$-dimensional dilaton and the $7$-dimensional space-time metric
and $2$-form  are singlets. The four vectors
obtained from the R-R forms transform 
in a chiral spinor representation and, after dualizing 
the $3$-form, the rest of the bosonic fields form a chiral spinor
of $2$-forms.
 
For our purpose, it will be enough to consider the
nonlinear sigma model part of the action containing the
kinetic terms  of the scalar fields of the theory
\[
{\cal S}=\, \frac{1}{2}\,\int d^{7}x~\sqrt{\widetilde{g}}\left( ~e^{-2\Phi}
~\widetilde{g}^{\mu \nu} ~{\rm tr}\,(
\partial_{\mu}{\cal M}^{-1}\partial_{\nu} {\cal M})+
\widetilde{g}^{\mu \nu} ~
\partial_{\mu} \chi^{T}    S({\cal M}) \,\partial_{\nu}\,\chi~\right)+\ldots,
\]
where $\widetilde{g}_{\mu \nu}$ and $\Phi$ are the $7$-dimensional metric and 
dilaton, and we have not written the kinetic term for the dilaton.
The nonlinear sigma model part of the action is written in a form that 
is explicitly  invariant 
under $SO(3,3)$ and in fact the whole supergravity action 
could be written in invariant form.
The duality transformations of the scalar fields are given by
\beqarr
\bar{\cal M}{}&=&\Lambda^{-T} {\cal M} \Lambda^{-1}, \nonumber  \\
\bar{\chi}&=& S(\Lambda) \chi. \label{chi}  
\eeqarr
To prove the invariance of the action we used $S(\Lambda^{T})=S(\Lambda)^{T}$.

The main purpose of this section was to obtain 
the relations~\rref{Cfields} which
show how the fields $\chi$ with simple transformations
properties under the T-duality group are related to the R-R forms.

\section{T-duality of the Chern-Simon Type Terms}
\label{DUAL}

In this section we show how to modify the NCSYM action so 
that it describes the DLCQ of M-theory in the presence of 
arbitrary moduli. In the auxiliary Type IIA string theory
the additional moduli are constant R-R backgrounds 
corresponding to  generalized Wilson lines.
Then we show that the action which includes the new terms 
is also invariant under the duality group $SO(3,3\,|{\bf Z})$ and that 
the parameters of the new terms transform exactly as 
expected from string theory.

First we guess the form of these terms using our experience with the
commutative case which corresponds to a vanishing NS-NS 
background $2$-form $B$. 
In this case the compactified Matrix model corresponding to $n$ D0-branes 
is described by a $U(n)$ supersymmetric Yang-Mills theory. This is obtained 
by performing a T-duality transformation 
along all the compact directions. However, for nonvanishing R-R
moduli, the action contains an additional Chern-Simon type term~\cite{ML} 
\[
{\cal S}_{CS} =
\frac{1}{4(2\pi)^{3}}
\int {\rm tr}\left(e^{2\pi {\cal F}} 
\sum_{k~{\rm odd}} A^{(k)}\right),
\]
where $A^{(k)}$ are the the T-dual R-R fields. Note also that
under T-duality in all directions the dual of $B$ also
vanishes if  $B$
was zero. This is why  only ${\cal F}$ appears in the exponent while
in general we would also subtract the dual of~$B$. 

Next we will consider the effect of a nonvanishing $B$ on this action.
If $\gamma^{ij}$ represents a two cycle wrapped around 
directions $x^{i}$ and $x^{j}$, then the deformation parameters
are defined by
\[
\Theta_{ij} = \frac{1}{(2\pi)^2} \int_{\gamma^{ij}}B.
\]
In the 
super Yang-Mills part of the action the only change 
required by a nonvanishing $B$
was to make the coordinates
noncommutative with deformation parameter $\Theta$.
The metric and gauge coupling constant are the same as those obtained by 
T-duality from the Matrix model for a vanishing NS-NS $2$-form.
We emphasize that the metric of the NCSYM gauge theory is not 
the  T-dual metric 
obtained by first taking the inverse of $E=G+\Theta$ and then
extracting the symmetric part.
The NCSYM metric  $G^{ij}$ is just the inverse of the original metric.
Thus we must distinguish between a T-duality in all directions and the
noncommutative Fourier transformation relating
the Matrix model and the NCSYM gauge theory. 

Let us explain why the 
NCSYM metric is $\Theta$ independent. To compactify the Matrix model 
on a torus we first consider the Matrix model on the covering space 
and then impose a quotient condition. If the  $B$ modulus is nonvanishing,
once we go to the topologically trivial 
covering space, we can 
gauge it away. However this gauge transformation does not leave the wave 
functions of strings invariant and thus we must transform the 
translation operators implementing the quotient condition. The new 
translation operators do not commute and their noncomutativity is 
measured by $\Theta$.

Imposing the new quotient conditions on the Matrix action
results directly in the NCSYM gauge theory. The only difference
with the $B=0$ case is that we have to use noncommutative Fourier 
transformations instead of the standard Fourier transformations
when we go from the Matrix model to the NCSYM gauge theory.
This however does not result in a
different metric and gauge coupling constant.
The main point of this discussion was to show that 
we can trade a nonvanishing $B$ field for noncommutative
coordinates on the dual super Yang-Mills gauge theory.

We will assume that the parameters of the  Chern-Simon terms are also the
same as for vanishing $\Theta$, except that the new terms are defined on a 
noncommutative torus. In particular for compactification on a three
torus we have
\beqn
{\cal S}_{CS}= \frac{1}{4(2\pi)^{3}}
\int {\rm tr} 
\left(2\pi
{\cal F} \wedge A^{(2)}+ \frac{1}{2}\, 2\pi {\cal F} \wedge 2\pi 
{\cal F}
\wedge A^{(0)}
\right).
\label{CSaction}
\eeqn
Just as in the commutative case these terms are topological,
supersymmetric and gauge invariant. 
In this action $\Theta$ only appears through the noncommutativity of the 
coordinates and $A^{(0)}$ and $A^{(2)}$ are the T-dual R-R 
forms\footnote{When we write the R-R forms in components 
we will drop the rank of the 
form as it is possible to identify the form from the
position and number of indices.} 
calculated as if 
the NS-NS $2$-form vanishes
\[
A^{(0)}= *  A_{(3)},~~
A^{(2)}= - *  A_{(1)}.
\]
The $1$-form R-R field $A_{(1)}$ has a lower index and should not 
be confused with the Yang-Mills gauge field $A^i$. 
With this distinction in mind 
we can write the action~\rref{CSaction} in the dual 
Matrix theory language using the R-R backgrounds on the original torus
\[
{\cal S}_{CS}= 
\int \,dt\, {\rm Tr} \left(
 \dot{X}^iA_i + \frac{i}{2\pi} \dot{X}^i X^j X^{k} A_{ijk}
\right),
\]
where ${\rm Tr}$ is the formal trace over infinite dimensional matrices
divided by the infinite order of the quotient group~\cite{WTc}.
It is convenient to write the action in component notation
\beqn
{\cal S}_{CS}= \frac{1}{2}
\int dt  \int \frac{d^3 \sigma}{(2\pi)^3} 
\, {\rm tr} \left(  \varepsilon_{ijk} 
\left(  2\pi
{\cal F}^{0i} A^{jk} + (2\pi)^{2}{\cal F}^{0i} {\cal F}^{jk}  
 A \right)\right),
\label{rraction}
\eeqn
where the magnetic and electric field strengths in the temporal gauge are
\[ 
{\cal F}^{0i} = i[\partial^{0},D^i], ~~
{\cal F}^{ij} = i[D^i,D^j].
\]

We now show that the action~(\ref{rraction}) is invariant  under 
the $\SOS$ duality group of the auxiliary string theory. 
Consider a Chern-Simon type action defined on a $\eta$-bundle.
Here $\eta$ is a $\SOS$ spinor containing the rank of the group and 
the magnetic  flux numbers
\[
\eta=
\left(
\begin{array}{c}
n  \\
M^{23}  \\
M^{31}  \\
M^{12} 
\end{array}
\right).
\]
We will perform the same sequence of 
field redefinitions used in~\cite{BMZ}, where it 
was shown explicitly for the case of vanishing R-R moduli, that
the  $U(n)$
NCSYM action is equivalent to a $U(q)$ NCSYM action on a trivial bundle, 
where $q$ is the greatest
common divisor of $n$ and the magnetic fluxes $M$. 
Let $H = ({\cal C}\Theta + {\cal D})^{-1}$ be the matrix defined 
in~\cite{BMZ}, where ${\cal C}$ and ${\cal D}$ are the lower 
block components of the $\SOS$ transformation relating the original 
 \mbox{NCSYM} gauge theory to the theory on the trivial bundle with $U(q)$ 
gauge group.  
Then we make the following
constant curvature connection and field
redefinitions
\[
\widehat{\nabla}^i \df (H^{-1})^{i}_{\,j} \nabla^j,
~\widehat{A}^i \df (H^{-1})^{i}_{\,j} A^j,
\]
\[
~\widehat{D}^i \df (H^{-1})^{i}_{\,j} D^j,
\]
\[
{\cal \widehat{F}}^{kl}=
[\widehat{\nabla}^k,\widehat{A}^l]-[\widehat{\nabla}^l,\widehat{A}^k]-
i[\widehat{A}^k,\widehat{A}^l].
\]
The curvature can be split into a constant term and a fluctuating piece
\beqn
{\cal F}^{ij} = {\cal F}_{(0)}^{ij} +
H^{i}_{\,k} H^{j}_{\,\,l}\,
{\cal \widehat{F}}^{kl},
\label{Fsubs}
\eeqn
\[
{\cal F}^{0k} = 
H^{k}_{\, \, l} \,
{\cal \widehat{F}}^{0l}.
\]
Using the matrices $Q$ and $R$ defined in~\cite{BMZ} 
we perform a change of integration variables 
$\widehat{\sigma}=\sigma QR$, which introduces a Jacobian factor
\beqn
\int d^{3}\sigma\, {\rm tr}\, \Psi(\sigma) =
\int  d^{3}\widehat{\sigma}\, \det( Q^{-1})\, 
{\rm tr}\, \Psi(\widehat{\sigma}(QR)^{-1}).  
\label{coordchange}
\eeqn

Making the substitutions~(\ref{Fsubs}),~(\ref{coordchange}) 
and collecting similar terms we find
\[
{\cal S}_{CS}=\frac{1}{2(2\pi)^2} 
\int dt  \int d^3 \widehat{\sigma} \, \frac{q}{n} \, 
{\rm tr} \,\left( \varepsilon_{ijk} 
\left(
{\cal \widehat{F}}^{0i} A'^{\,jk} +  2\pi 
{\cal \widehat{F}}^{0i}{\cal \widehat{F}}^{jk}  
 A' \right)\right),
\]
where
\[
q A' =  (n \det{Q^{-1}} \det{H})\, A,
%\label{uglyvartheta}
\]
\[
\varepsilon_{ijk} \,q A'^{\,ij}
= H^l_{\, k}  \,\varepsilon_{ijl} \, n \det{Q^{-1}}
\,(A^{\,ij} + 2\pi {\cal F}_{(0)}^{ij} A) .
%\label{uglyCij}
\]
One can now rewrite the action in terms of new operators $\sigma'_i$, 
$\partial'^i$, and $U'$, and a $q$ dimensional trace. 
See~\cite{BMZ} for a more detailed discussion of this
substitution.
\[
{\cal S}_{CS}= \frac{1}{4(2\pi)^{3}}
\int {\rm tr}_{q} 
\left(2\pi
{\cal F}' \wedge A'^{(2)}+ \frac{1}{2}\, 2\pi {\cal F}' \wedge 2\pi 
{\cal F}'
\wedge A'^{(0)}
\right).
\]

More generally  the action is invariant under duality transformations 
if the Chern-Simon parameters are related as follows
\beqn
~~~~~~~~\bar{A}_{(0)} = |\det \CTPD |^{-1/2} A_{(0)},\label{A0trans}
\eeqn
\[
*(\bar{A}^{(2)} + 2\pi \bar{{\cal F}}_{(0)} \bar{A}^{(0)}) 
\bar{n} \det \bar{Q}^{-1}
= \CTPD^{-T}
*(A^{(2)} + 2\pi {\cal F}_{(0)} A^{(0)}) 
n \det Q^{-1} ,
\]
where ${\cal C}$ and ${\cal D}$ are the lower block components of the $\SOS$ 
matrix relating 
the two theories, and the star operator 
is the duality operator acting only with respect
to the compact coordinates.

Next we write the Chern-Simon parameters in term of the fields $C$
discussed in Section~\ref{SUGRA}
\beqn
A^{(0)} =  *( C_{(3)} +
C_{(1)}\wedge \Theta),
\label{defvartheta}
\eeqn
\beqn
A^{(2)} = - * C_{(1)} \label{defCij}.
\eeqn
%and using the identities~\rref{use2} listed in the Appendix
% we can rewrite these 
%relations in terms of the $C$ moduli. 
To obtain a  compact form first
define the column matrices $u$ and $v$ with components
\beqarr
u_i &=& n C_i -\frac{1}{2} M^{jk} C_{ijk}, \nonumber \\
v^i &=& M^{ij}C_j. \nonumber
\eeqarr
If $\chi$ transforms as a spinor, $u$ and $v$ are the block components of
a $\SOS$ vector as shown in the Appendix~\rref{goodreason}.
Then using the identities~\rref{use2}  listed in the Appendix,
the transformation~\rref{A0trans} can be written as
\beqn
\bar{C}_{(3)} +\bar{C}_{(1)}\wedge \bar{\Theta}=
|\det \CTPD |^{-1/2}  \,
(C_{(3)} +C_{(1)}\wedge \Theta),
\label{Str}
\eeqn
\beqn
(\bar{u}-\bar{\Theta}\bar{v})
=\CTPD^{-T}(u-\Theta v).
\label{Vtr}
\eeqn
Comparing~\rref{Str} and~\rref{Vtr} with~\rref{Strans} and~\rref{Vtrans} in 
the Appendix we see that the R-R fields must transform in a spinor
representation of $\SOS$
\[
\bar{\chi}=S(\Lambda) \chi.
\] 

Thus the duality transformations of all the 
parameters of the NCSYM, including those of the  Chern-Simon type terms,  
coincide with the transformation of moduli of the Type IIA strings 
compactified on a torus in the limit of vanishing $\alpha'$ and $G_{ij}$.

Using the transformation properties of $g^{-2}_{SYM}$ and $A^{(0)}$
it follows that the complex coupling 
\[
\tau= A^{(0)}+\frac{4\pi i}{g^{2}_{SYM}}
\]
also transforms
simply under the T-duality group with the same $\Theta$ dependent factor appearing 
in~\rref{A0trans}. 

Finally note that the BPS spectrum corresponding to the electric fluxes 
obtained in~\cite{HV,DBBM,KS} is modified in the presence of nonvanishing R-R
moduli 
\[
{\cal E}^{U(n)} =\frac{g_S}{2}\,
\, | n - \frac{1}{2} {\rm tr}(M \Theta)|^{-1}
\]
\beqn
\times
((n_i-u_i)-\Theta_{ik} (m^k-v^k))\,
G^{ij} \,
((n_j-u_j)- \Theta_{jl} (m^l-v^l)),
\eeqn
where we used the notation of~\cite{DBBM}.

\section*{Acknowledgments}
This work was supported in part by 
the Director, Office of Energy Research, Office of High Energy and Nuclear
Physics, Division of High Energy Physics of the U.S. Department of Energy
under Contract DE-AC03-76SF00098 and in part by the National Science 
Foundation under grant PHY-95-14797.

\appendix
\section*{Appendix}

In this Appendix we present some mathematical results regarding the 
spinor representation of the T-duality group and duality invariant
quantities in the small volume limit.
The group $SO(d,d)$ is the group of $2d$-dimensional 
matrices $\Lambda$ satisfying  $\Lambda J\Lambda^{T}=J$ where $J$\, 
is a matrix 
with the block form
\[
J=
\left(
\begin{array}{cc}
0 & 1 \\
1 & 0
\end{array}
\right).
\]
It will be useful to know how to calculate the
Weyl spinor representation matrix of an $SO(3,3)$
group element $\Lambda$ with the block form
\beqn
\Lambda=
\left(
\matcc
{\cal A} & {\cal B}  \\
{\cal C} & {\cal D} 
\emat
\right).
\]
First note that if ${\cal A}$ is invertible $\Lambda$ has a block 
Gauss decomposition
\[
\Lambda = 
\left(
\begin{array}{cc}
1 & 0 \\
{\cal C}{\cal A}^{-1} & 1
\end{array}
\right)\left(
\begin{array}{cc}
{\cal A} & 0 \\
0 & {\cal A}^{-T}
\end{array}
\right)\left(
\begin{array}{cc}
1 & {\cal A}^{-1} {\cal B} \\
0 & 1
\end{array}
\right),
\label{Decomp}
\eeqn
where one can show using the group relations that ${\cal C}{\cal A}^{-1}$ 
and ${\cal A}^{-1} {\cal B}$ are antisymmetric.
This decomposition is in fact true for 
generic $SO(d,d)$ matrices.
For $d=3$ we can give the explicit spinor representation matrices 
for each factor
in~\rref{Decomp} thus obtaining the spinor representation of a 
generic $SO(3,3)$ matrix $\Lambda$
\beqn
S  = 
\left(
\begin{array}{cc}
1 & 0 \\  
{*}({\cal C}{\cal A}^{-1}) & 1
\end{array}
\right)\left(
\begin{array}{cc}
\det {\cal A}^{1/2} & 0 \\
0 & \det {\cal A}^{-1/2} {\cal A}
\end{array}
\right)\left(
\begin{array}{cc}
1 & *({\cal A}^{-1} {\cal B})^{T}  \\
0 & 1
\end{array}
\right).
\label{SofL}
\eeqn
The star denotes the duality operator. When acting on antisymmetric 
$3$-dimensional square matrices it gives the dual column matrix.

We can form invariants using two column matrices transforming in the vector
representation of $\SOdd$ and the symmetric $\SOdd$ matrix~${\cal M}$
\[
(r^{T},s^{T}) 
 {\cal M}
\left(
\matc
u  \\
v
\emat
\right).
\]
In the limit when $G$ goes to zero, using the block Gauss
decomposition of ${\cal M}$
\beqn
{\cal M} = 
\left(
\begin{array}{cc}
1 & 0 \\
B & 1
\end{array}
\right)\left(
\begin{array}{cc}
G^{-1} & 0 \\
0 & G
\end{array}
\right)\left(
\begin{array}{cc}
1 & -B \\
0 & 1
\end{array}
\right),
\label{decompo}
\eeqn
and after identifying $B$ with $\Theta$ we obtain the following 
invariant\footnote{To obtain a finite result, one should insert appropriate
factors of 
$\alpha'$ in~\rref{decompo} and also take $\alpha'$ to zero as 
discussed in the Introduction.}
\[
(r-\Theta s)^{T} G^{-1} (u-\Theta v).
\]
Using the transformation of $G$ under the duality 
group~\rref{dual} 
we can write the transformation of $u-\Theta v$
\beqn
(\bar{u}-\bar{\Theta}\bar{v})
=\CTPD^{-T}(u-\Theta v).
\label{Vtrans}
\eeqn

The spinor representation matrix of ${\cal M}$  
can be calculated using~\rref{SofL}
\[
S({\cal M}) = 
\left(
\begin{array}{cc}
1 & 0 \\  
b  & 1
\end{array}
\right)\left(
\begin{array}{cc}
\det G^{-1/2} & 0 \\
0 & \det G^{1/2} G^{-1}
\end{array}
\right)\left(
\begin{array}{cc}
1 & b^T  \\
0 & 1
\end{array}
\right).
\]
Then we can also form the  invariants ${\eta}^{T} S({\cal M}) \chi$ using 
two chiral spinors
\[
\chi=
\left(
\matc
\chi_{0} \\
\chi_{1} \\
\chi_{2} \\
\chi_{3}
\emat
\right),~~
\eta=
\left(
\matc
\eta_{0} \\
\eta_{1} \\
\eta_{2} \\
\eta_{3}
\emat
\right).
\]
In the limit of vanishing $G$, the invariant 
becomes
\beqn
(\eta_{0}+\frac{1}{2}\varepsilon^{ijk}\Theta_{ij}\eta_{k})
\,\sqrt{\det G^{-1}}\,
(\chi_{0}+\frac{1}{2}\varepsilon^{ijk}\Theta_{ij}\chi_{k}).
\label{Sinv}
\eeqn
From~\rref{Sinv} we obtain the following transformation law 
\beqn
\bar{\chi}_{0}+\frac{1}{2}\varepsilon^{ijk}\bar{\Theta}_{ij}\bar{\chi}_{k} =
|\det\, \CTPD |^{-1/2}\,
(\chi_{0}+\frac{1}{2}\varepsilon^{ijk}\Theta_{ij}\chi_{k}).
\label{Strans}
\eeqn
One can also check the relations~\rref{Vtrans} and~\rref{Strans} directly using the 
trasformations~\rref{Theta299} of
$\Theta$.

Given two chiral spinors $\chi$ and $\eta$ we first 
write them as Dirac spinors
\[
\chi_{D}= 
\left(
\matc
\chi \\
0
\emat
\right),~~
\eta_{D}= 
\left(
\matc
\eta \\
0
\emat
\right).
\]
Then using the same definition for $a^i$ and $a^{\dagger}_i$ as in the 
Appendix of~\cite{BMZ} we can form the $\SOS$ vector
\beqn
\left(
\matc
u_i  \\
v^i
\emat
\right)
=
\bar{\eta}_{D}
\left(
\matc
a^{\dagger}_i  \\
a^i
\emat
\right)
\chi_D.
\label{Vspinor}
\eeqn
where $\bar{\eta}_{D}=\eta^{\dagger} T$. Here $T$ is a matrix acting on 
Dirac spinors and plays the same role as $\gamma_0$ when we form
barred spinors in Minkowski space. It is given by
\[
T= (a^{\dagger}_1+a^1) (a^{\dagger}_2+a^2) (a^{\dagger}_3+a^3).
\]
Writing out all the spinor components in~\rref{Vspinor}
we have
\beqn
\left(
\matc
u_i  \\
v^i
\emat
\right)
=
\left(
\matc
 \eta_i\chi_0 - \eta_0 \chi_i  \\
\varepsilon^{ijk} \eta_j \chi_k
\emat
\right).
\label{goodreason}
\eeqn
Then  $u-\Theta v$ transforms as in~\rref{Vtrans} under the duality group. 
Such an expression, involving two chiral spinors and $\Theta$, 
is used in the main text.

Finally we list some useful identities presented 
in~\cite{BMZ}  
\beqarr
\det{H} &=& \left(q  \det (Q) /n  \right)^2,
\label{use2}
\\ 
M &=&2\pi n  Q^{-1}{\cal F}_{(0)} \, Q^{-T},
%\label{use3}
\\
 M^{ij} \varepsilon_{ijl} &=&
M^{ij} \varepsilon_{ijk}Q^{k}_{\, l}.
%\label{use4}
\eeqarr

\end{document}